\documentclass[aps,prl,twocolumn, preprintnumbers,superscriptaddress, showpacs]{revtex4}

\usepackage{graphicx}
\usepackage{dcolumn}

\begin{document}

\preprint{Ver. 4wsm}

\title{Complete Fermi surface in BaFe$_2$As$_2$ observed via Shubnikov-de Haas oscillation measurements on detwinned single crystals}


\author{Taichi Terashima}
\author{Nobuyuki Kurita}
\affiliation{National Institute for Materials Science, Tsukuba, Ibaraki 305-0003, Japan}
\affiliation{JST, Transformative Research-Project on Iron Pnictides (TRIP), Chiyoda, Tokyo 102-0075, Japan}
\author{Megumi Tomita}
\affiliation{National Institute for Materials Science, Tsukuba, Ibaraki 305-0003, Japan}
\author{Kunihiro Kihou}
\author{Chul-Ho Lee}
\author{Yasuhide Tomioka}
\author{Toshimitsu Ito}
\author{Akira Iyo}
\author{Hiroshi Eisaki}
\affiliation{JST, Transformative Research-Project on Iron Pnictides (TRIP), Chiyoda, Tokyo 102-0075, Japan}
\affiliation{National Institute of Advanced Industrial Science and Technology (AIST), Tsukuba, Ibaraki 305-8568, Japan}
\author{Tian Liang}
\author{Masamichi Nakajima}
\author{Shigeyuki Ishida}
\author{Shin-ichi Uchida}
\affiliation{JST, Transformative Research-Project on Iron Pnictides (TRIP), Chiyoda, Tokyo 102-0075, Japan}
\affiliation{Department of Physics, University of Tokyo, Bunkyo-ku, Tokyo 113-0033, Japan}
\author{Hisatomo Harima}
\affiliation{JST, Transformative Research-Project on Iron Pnictides (TRIP), Chiyoda, Tokyo 102-0075, Japan}
\affiliation{Department of Physics, Graduate School of Science, Kobe University, Kobe, Hyogo 657-8501, Japan}
\author{Shinya Uji}
\affiliation{National Institute for Materials Science, Tsukuba, Ibaraki 305-0003, Japan}
\affiliation{JST, Transformative Research-Project on Iron Pnictides (TRIP), Chiyoda, Tokyo 102-0075, Japan}


\date{\today}

\begin{abstract}
We show that the Fermi surface (FS) in the antiferromagnetic phase of BaFe$_2$As$_2$ is composed of one hole and two electron pockets, all of which are three dimensional and closed, in sharp contrast to the FS observed by angle-resolved photoemission spectroscopy.  Considerations on the carrier compensation and Sommerfeld coefficient rule out existence of unobserved FS pockets of significant sizes.  A standard band structure  calculation reasonably accounts for the observed FS, despite the overestimated ordered moment.  The mass enhancement, the ratio of the effective mass to the band mass, is 2--3.
\end{abstract}


\maketitle



\newcommand{\ud}{\mathrm{d}}
\def\degree{\kern-.2em\r{}\kern-.3em}

Precise knowledge of electronic structures near the Fermi level $E_F$ is a prerequisite for understanding the origins of novel superconductivity.  In the case of the iron-pnictide superconductors\cite{Kamihara08JACS}, ordinary electronic band structure calculations of the parent compounds greatly overestimate the antiferromagnetic (AFM) ordered moment\cite{Yin08PRL, Mazin09NatPhys, Ishibashi08JPSJS}.  Several moment-reduction methods have been proposed, e.g., pnictogen-height adjustment\cite{Mazin09NatPhys}, negative-$U$\cite{Nakamura08condmat}.  Moreover, a dynamical mean-field theory (DMFT) study suggests that correlations seriously modify the Fermi surface (FS) in the AFM phase\cite{Yin11NatPhys}.  There is a clear need for an experimentally determined benchmark FS for judging how successfully those theoretical approaches reproduce band structures.



Recently, it was shown that BaFe$_2$As$_2$ can be detwinned mechanically by compressing or elongating single crystals along a tetragonal [110] axis\cite{Chu10Science, Tanatar10PRB}.  This allows one to resolve intrinsic in-plane anisotropy in the AFM phase.  In this Letter, we completely determine the FS in the AFM phase of BaFe$_2$As$_2$ via Shubnikov-de Haas (SdH) oscillation measurements on detwinned single crystals.  The determined FS is reasonably accounted for by our band structure calculation but appreciably differs from the FS observed by previous angle-resolved photoemission spectroscopy (ARPES) studies \cite{LiuC09PRL, Yang09PRL, Malaeb09JPSJ, Richard10PRL, Yi11PNAS}.  We also consider implications of the present results for transport properties.

High-quality BaFe$_2$As$_2$ single crystals with the residual resistivity ratios of 40--60 were prepared by a self-flux method and subsequent anneal\cite{Nakajima11PNAS}.  A device similar to one described in Ref.~\onlinecite{Nakajima11PNAS} was used to mechanically detwin them.  Standard four-contact resistivity $\rho$ measurements were performed in a dilution refrigerator and superconducting magnet.  For a sample compressed (elongated) along a tetragonal [110] axis, the electrical current was applied parallel to the compression (elongation) direction, resulting in $I \parallel b$ ($I \parallel a$) geometry [Fig. 1(b)].  The electronic band structure of BaFe$_2$As$_2$ was calculated within the local spin-density approximation (LSDA) using a full-potential LAPW method (TSPACE and KANSAI-06).  The experimental crystal structure at $T$ = 20 K \cite{Rotter08PRB} was used.  The AFM order of the Fe moments was incorporated using the space group $Cccm$.  The calculated magnetic moment is 1.6 $\mu_B$/Fe.

\begin{figure*}
\includegraphics[width=16cm]{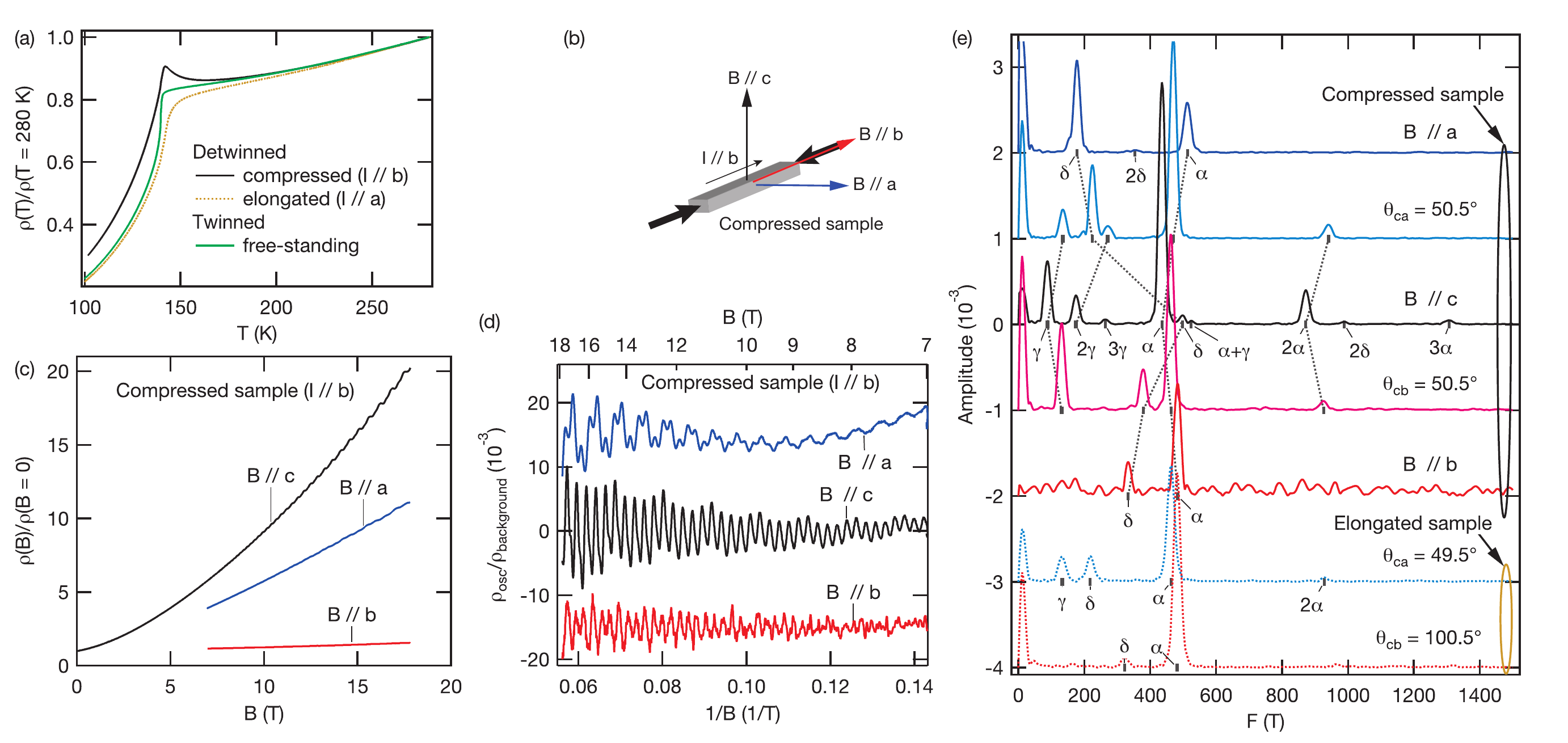}
\caption{\label{fig1}(color online).  (a) $T$-dependence of $\rho$ near the structural/magnetic phase transition.  (b) Schematic diagram of the compressed sample.  (c) $B$-dependence of $\rho$ in the compressed sample for three principal field directions at $T \sim$0.17 K.  The small magnetoresistance for $B \parallel b$ results from the longitudinal configuration $I \parallel B$.  (d) SdH oscillations extracted from the data in (c) as a function of inverse field $1/B$.  The curves are vertically shifted for clarity.  A third polynomial was fitted to each $\rho(B)$ curve to determine a smoothly varying background $\rho_{background}$ and was subtracted to obtain an oscillatory part $\rho_{osc}$.  A normalized quantity $\rho_{osc}/\rho_{background}$ is shown.  (e) Fourier transforms of the SdH oscillations in the compressed and elongated samples for selected field directions.  The field angle $\theta_{ca}$ ($\theta_{cb}$) is that between the $c$ axis and $B$ in the $ca$ ($cb$) plane.  The spectra are vertically shifted for clarity.  The two samples give consistent results (compare the elongated-sample spectra at $\theta_{ca}$ = 49.5$^{\circ}$ and  $\theta_{cb}$ = 100.5$^{\circ}$ with the compressed-sample spectra at $\theta_{ca}$ = 50.5$^{\circ}$ and $B \parallel b$, respectively).}   
\end{figure*}

\begin{figure*}
\includegraphics[width=16 cm]{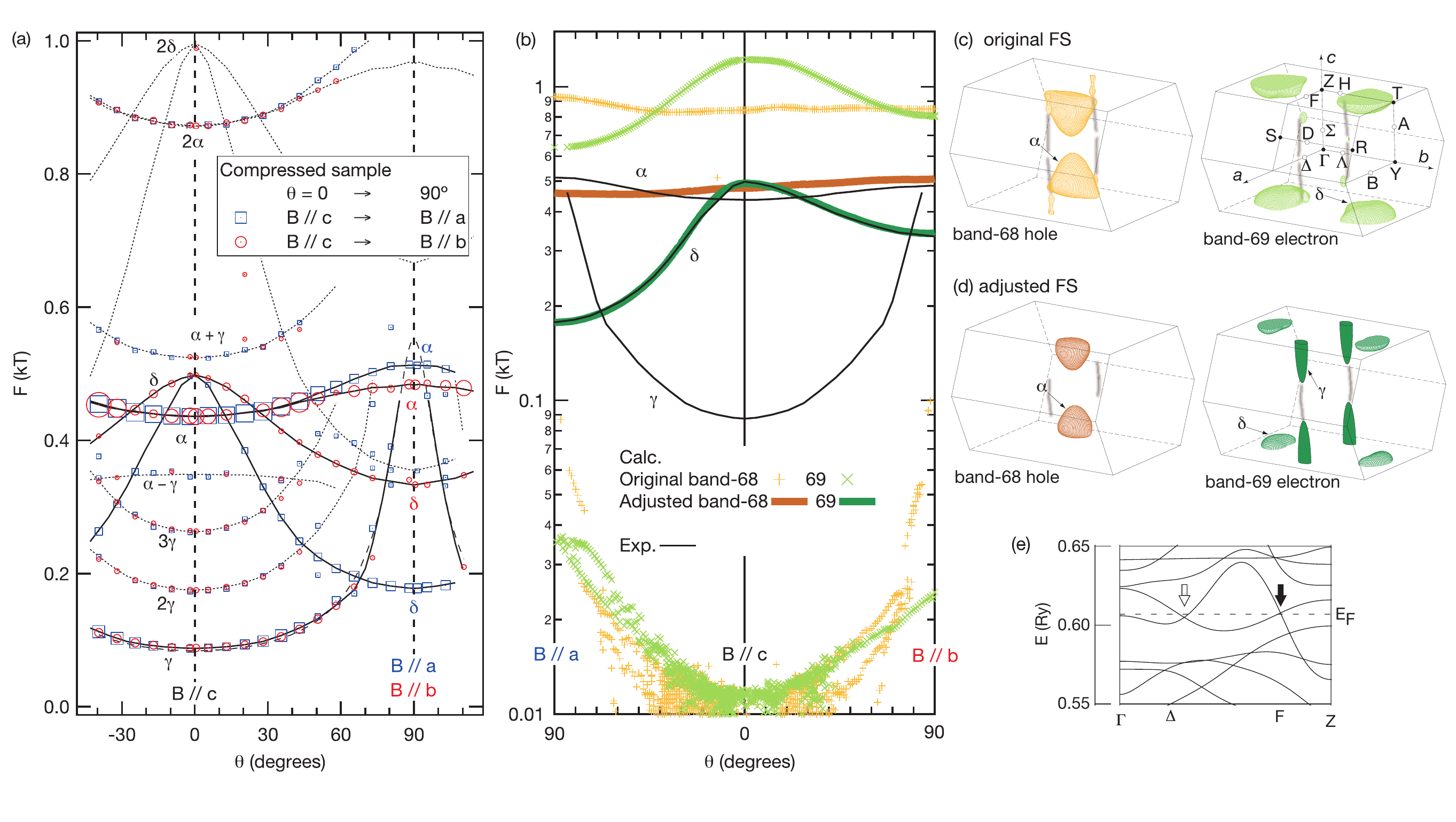}
\caption{\label{fig1}(color online)  (a) Angle dependence of the experimental SdH frequencies in the compressed sample.  The mark sizes indicate the oscillation amplitudes.  The $ca$-plane (blue squares) and $cb$-plane (red circles) data are superimposed.  The solid lines indicate the fundamental frequency branches.  The dotted lines indicating the harmonic and combination frequencies are calculated from the solid lines.  The dashed line, almost overlapped by a solid line, is a fit to the $\gamma$ branch assuming that the responsible FS pocket is an ellipsoid of revolution.  The fit gives $k_F^{ab}$ = 0.050 \AA$^{-1}$ and $k_F^c$ = 0.33 \AA$^{-1}$, $k_F^{ab (c)}$ being the Fermi wave number in the $ab$ plane (along the $c$ axis).  No frequency other than 3$\alpha$ is observed above $F$ = 1 kT.  (b) Comparison between the calculated and experimental frequencies.  For the calculated frequencies, both original and adjusted ones are shown (see text).  The solid curves representing the experimental frequencies are the same as those in (a).  (c) The FS resulting from the original band structure calculation.  It consists of a band-68 hole sheet (left) and band-69 electron sheet (right).  Since our band structure calculation does not include spin-orbit coupling, it predicts chains of alternating hole and electron pockets along the grey lines, in accord with Ref.~\onlinecite{Ran09PRB}, but they are so thin that they can not be drawn accurately.  The directions $a$, $b$, and $c$ shown in the figure refer to those of the $Fmmm$ orthorhombic unit cell\cite{Rotter08PRB}.  (d) The FS in BaFe$_2$As$_2$ determined in the present study.  The drawings of the $\alpha$ hole and $\delta$ electron pockets are based on the adjusted band structure calculation, and the $\gamma$ electron pocket is schematically shown, based on the ellipsoid fit in (a).  There might be minute FS pockets along the grey lines, though our data show no trace of such pockets.  (e) Electronic band structure near $E_F$ along $\Gamma\Delta$FZ.  The $\gamma$ pocket is placed at the position marked by the black arrow (see text).}   
\end{figure*}

\begingroup
\squeezetable
\begin{table}
\caption{\label{Tab1} $m^*$ and $\tau$.  $m_e$ is the free electron mass.}
\begin{ruledtabular}
\begin{tabular}{ccddd}
Fermi & Field & & &  \\
surface & direction & \multicolumn{1}{c}{$m^*/m_e$} & \multicolumn{1}{c}{$m^*/m_{band}$}  & \multicolumn{1}{c}{$\tau$ (10$^{-12}$ s)}\\
\hline

\multicolumn{1}{c}{$\alpha$ hole} & \multicolumn{1}{c}{$B \parallel a$} & 2.8(1) & 3.0 & 1.4(1)\\
 & \multicolumn{1}{c}{$B \parallel b$} & 2.7(1) & 3.6 & 2.8(4)\\
 & \multicolumn{1}{c}{$B \parallel c$} & 2.1(1) & 2.8 & 2.3(2)\\
\\

\multicolumn{1}{c}{$\gamma$ electron} & \multicolumn{1}{c}{$B \parallel c$} & 0.9(1) &  & 2.6(10)\\
\\

\multicolumn{1}{c}{$\delta$ electron} & \multicolumn{1}{c}{$B \parallel a$} & 1.3(1) & 2.0 & 1.0(2)\\
 & \multicolumn{1}{c}{$B \parallel b$} & 2.1(1) & 3.0 & 2.0(4)\\
 & \multicolumn{1}{c}{$B \parallel c$} & 2.4(3) & 1.7 & \\
\end{tabular}
\end{ruledtabular}
\end{table}
\endgroup

Figure 1(a) shows $\rho$ as a function of temperature $T$ near the structural/magnetic phase transition for compressed, elongated, and freestanding (= twinned) samples.  The $\rho(T)$ curves are consistent with previous data\cite{Chu10Science, Tanatar10PRB}.  Figure 1(c) shows the magnetic-field $B$ dependence of $\rho$ in the compressed sample [Fig. 1(b)] for $B \parallel a$, $b$, and $c$ at $T\sim0.17$ K.    After subtracting the smooth background, SdH oscillations are clearly visible for all field directions [Fig. 1(d)].  The oscillations continue to below $B$ = 7 T, much lower than previously reported\cite{Sebastian08JPCM, Analytis09PRB}.  Figure 1(e) shows Fourier transforms of SdH oscillations in the compressed and elongated samples for selected field directions.  The two samples gave consistent results.  Figure 2(a) summarizes the angular dependence of the SdH oscillations in the compressed sample.  Clear differences are seen between the $ca$ (blue squares) and $cb$ (red circles) planes, confirming appropriate detwinning.

We find three fundamental frequency branches $\alpha$, $\gamma$, and $\delta$ in Fig. 2(a).  The first two branches are consistent with (hence named as in) the previous reports\cite{Sebastian08JPCM, Analytis09PRB}, but the last branch $\delta$ has not been observed previously.  We attribute the previously reported $\beta$ branch to the second harmonic of $\gamma$ since its angular dependence agrees well with the angular dependence of $\gamma$.  This assignment is further supported by the observation that $m^*$ associated with this frequency is twice that of $\gamma$.  The $\alpha$ frequency shows only small frequency anisotropy, $F$ for $B \parallel a$ being slightly larger than that for $B \parallel b$, but its amplitude shows clear anisotropy between the $ca$ and $cb$ planes (note the mark sizes in Fig. 2a), which might be explained by anisotropic spin-splitting effects\cite{SM}.  The $\gamma$ branch shows no detectable in-plane anisotropy.  The angular variation fits that expected for an FS pocket whose shape is a prolate ellipsoid of revolution, as indicated by the dashed line.  The angle dependence of the $\delta$ branch indicates that the responsible FS pocket is flattened along the $c$ axis and exhibits the most pronounced frequency anisotropy between the $ca$ and $cb$ planes.  The effective mass $m^*$ and electron relaxation time $\tau$ were determined from the $T$- and $B$-dependences of the SdH oscillation amplitudes, respectively, as usual\cite{Shoenberg84}, and are tabulated in Table I.

Our original band structure calculation indicates that bands-68 and -69 cross $E_F$, giving hole and electron sheets of the FS, respectively [Fig. 2(c)].  Each sheet consists of large and small pockets.  The frequencies originating from the large hole and electron pockets are in the region of 1 kT, and those from the small pockets are in the region of 0.01 kT for $B \parallel c$ [pale-colored + and x in Fig. 2(b)].  Although the experimental and calculated frequencies differ by a factor of approximately two, it is clear from the angular dependence that the observed $\alpha$ and $\delta$ branches can be assigned to the large hole and electron pockets, respectively, of the calculated FS.

To improve agreement between the experimental and calculated $\alpha$ and $\delta$ frequencies, we shifted the energies of bands-68 and -69 by -0.0032 Ry (-44 meV) and +0.0048 Ry (+65 meV), respectively [Fig. 2(b) thick curves and Fig. 2(d)].  Similar procedures have been used in other multiband metals very successfully\cite{Carrington05PRB, Analytis09PRL}.  The agreement now is satisfactory (our criterion for $\alpha$ is the average of the frequencies for the three principal directions), though the observed small angular dependence of $\alpha$ is not reproduced so well as in the original calculation.  The adjusted bands-68 and -69 contain 0.0235 holes and 0.0130 electrons per primitive cell, respectively [the primitive cell contains two formula units (fu)].  The densities of states (DOS) at $E_F$ are 6.64 and 6.35 states/Ry per primitive cell for the adjusted bands-68 and -69, respectively.  Comparing the measured effective masses with the band masses $m_{band}$ (Table I), we find average mass enhancements $m^*/m_{band}$ of 3.1 and 2.3 for the band-68 $\alpha$ and band-69 $\delta$ pockets, respectively; these are small compared to the enhancements of 3--20 found in KFe$_2$As$_2$\cite{Terashima10JPSJ}.  Using these values together with the calculated DOS, we can estimate that bands-68 and -69 contribute 1.8 and 1.2 mJ/K$^2$mol-fu, respectively, to the Sommerfeld coefficient of the specific heat.

We now consider the $\gamma$ pocket.  It is reasonable to assume that the $\gamma$ pocket is located at a position where a small pocket appeared in the original calculation.  Figure 2(e) shows the relevant part of the band structure; two likely positions are marked by arrows.  The band crossings at these points, sometimes referred to as Dirac nodes\cite{Ran09PRB, Richard10PRL}, become anticrossings if spin-orbit coupling, which is absent in our calculation, is included and hence do not necessarily give FS pockets: $E_F$ may occur in a gap.  A recent laser ARPES study\cite{Shimojima10PRL}, which is considered bulk sensitive because of the low photon energy of 7 eV, has however shown that the right (anti)crossing marked by the black arrow actually occurs below $E_F$ and produces an electron pocket of non-negligible size.  We identify this pocket with our $\gamma$ pocket.  The (anti)crossing is situated almost at $E_F$ in the original calculation, and the renormalized Fermi energy for the $\gamma$ pocket is estimated to be 11 meV from the aforementioned ellipsoid fit and measured effective mass.  Assuming the mass enhancement to be about three, a band shift of $\sim$30 meV is enough to produce the $\gamma$ pocket.  Because of the symmetry, two $\gamma$ pockets occur in the BZ [Fig. 2(d)] and enclose 0.0116 electrons, resulting, within error, in perfect carrier compensation (0.0235 holes vs. 0.0246 electrons).  Using the measured effective mass for $\gamma$, we estimate the contribution to the Sommerfeld coefficient to be 1.9 mJ/K$^2$mol-fu.  The sum of the contributions from the $\alpha$, $\gamma$, and $\delta$ pockets is 5.0 mJ/K$^2$ mol-fu, which is in excellent agreement with a direct measurement on an annealed single crystal (5.1mJ/K$^2$ mol-fu)\cite{Rotundu10PRB}.  The carrier compensation and agreement on the Sommerfeld coefficient rule out existence of unobserved pockets with a comparable volume to the observed ones.  Although there might be minute pockets with at least an order-of-magnitude smaller volume along the grey lines in Fig. 2(d), we have seen no trace of such pockets.

Our band structure calculation accounts well for the observed FS if band energy adjustments of at most 65 meV are allowed, despite the fact that the calculated magnetic moment 1.6 $\mu_B$ overestimates the experimental one 0.87 $\mu_B$\cite{Huang08PRL}.  The magnitudes of the adjustments are similar to those necessary in a prototypical multiband metal MgB$_2$ (up to $\sim 90$meV)\cite{Carrington05PRB}.  Since our calculation is based on the experimental crystal structure and does not include ad hoc procedures such as the As-position adjustment and negative-$U$, those procedures are unjustified.  The observed FS is very different from the FS of the DMFT study of Ref.~\onlinecite{Yin11NatPhys}, providing a strong argument against the claim that electronic correlations largely modify the FS of the AFM phase.

The agreement between our FS and those observed in previous ARPES measurements\cite{LiuC09PRL, Yang09PRL, Malaeb09JPSJ, Richard10PRL, Yi11PNAS} is limited.  For example, the $\delta$ bright spot, the $\epsilon$ petal electron pocket, and one of the $\alpha$ and $\beta$ hole pockets reported in Ref.~\onlinecite{Yi11PNAS} may correspond to our $\gamma$, $\delta$, and $\alpha$ pockets, respectively, but the other reported pockets are absent in our FS.  Further, warped FS cylinders often found in $k_z$-resolved ARPES measurements\cite{LiuC09PRL, Malaeb09JPSJ, Yi11PNAS} are incompatible with our FS composed of closed pockets.  It should be noted that the photoelectron escape depth ($\sim$ 5 \AA) for a typical photon energy of 20$\sim$40 eV\cite{Seah79SIA} is shorter than the $c$-axis length (13 \AA), which inevitably limits the $k_z$ resolution.

Our data do not support recent transport studies\cite{Huynh11PRL} attributing observation of the linear-in-$B$ magnetoresistance to Dirac fermion transport in the quantum limit.  Our $\rho(B)$ curve for $B \parallel c$ is clearly superlinear [Fig. 1(c)].  The $\gamma$ pocket, a supposed Dirac pocket, does not reach the quantum limit until $B$ becomes comparable to $F_{\gamma}$ = 90 T.  The conductivities of the $\alpha$ and $\gamma$ pockets are estimated from $m^*$ and $\tau$ ($B \parallel c$) in Table 1 to be 3.6(2) and 5(2) $\times 10^6$ m$^{-1}$$\Omega^{-1}$, respectively.  The sum is already comparable to the measured conductivity, 11(2) $\times 10^6$ m$^{-1}$$\Omega^{-1}$.  It is therefore also unlikely that unobserved Dirac pockets, if any, dominate the conduction.

The origin of the in-plane resistivity anisotropy, which increases up to $\rho_b/\rho_a \sim$2 with only a few \% Co-doping\cite{Chu10Science, Tanatar10PRB}, is left elusive.  The in-plane mass anisotropy is negligible for the $\alpha$ pocket and opposite to the resistivity anisotropy for the $\delta$ pocket, i.e., $m^*$ along the $b$ axis being smaller than $m^*$ along the $a$ axis [Table 1, note $m^*$ for $B \parallel a (b)$ is associated with cyclotron motion in the $bc (ac)$ plane].  Intriguingly, this does not seem compatible with an optical study\cite{Dusza11EPL}, which indicates that the Drude spectral weight ($\omega_p^2=ne^2/\epsilon_0 m$ in a free-electron model) for $E \parallel a$ increases and becomes much larger than that for $E \parallel b$ below $T_N$.  It might be better reconciled with a more recent study\cite{Nakajima11PNAS}  suggesting an isotropic Drude component.

\begin{acknowledgments}
TT thanks Takahiro Shimojima, Tamio Oguchi, and Hiroaki Ikeda for helpful discussions.
\end{acknowledgments}

\newpage
\begin{figure}
{\Large Supplemental material} 
\includegraphics[width=8cm]{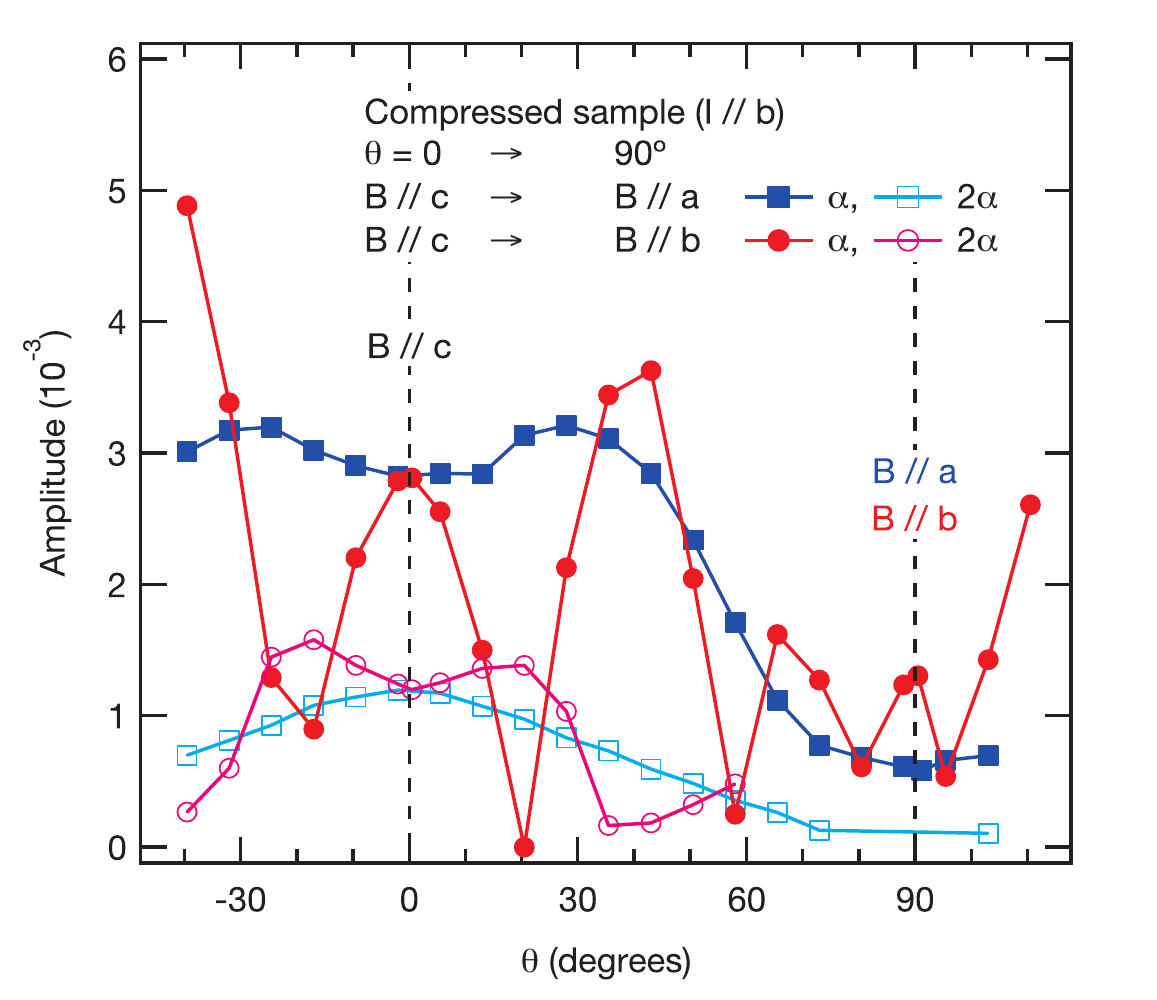}
\end{figure}
\noindent FIG.  Angle dependence of the $\alpha$ and 2$\alpha$ oscillation amplitudes.  In the $cb$ plane (filled red marks) the amplitude of $\alpha$ is suppressed near $\theta = \pm 20$ and $60^{\circ}$, but no such suppression is observed in the $ca$ plane (filled blue marks).  The observed suppression might be attributed to spin-zeros.$^1$  The measured quantum oscillation is a superposition of the oscillations coming from the spin-up and the spin-down electrons.  They have the same frequency but a different phase, resulting in interference of the two oscillations.  The net oscillation amplitude is therefore the single-spin oscillation amplitude multiplied by the spin-splitting factor $R_{s}=\cos(p \pi S)$, where $p$ is a harmonic number and $S$ = $(1/2)g_{eff}(m^*/m_e)$.  The effective $g$-factor $g_{eff}$ has to appropriately incorporate effects of many-body renormalization, spin-orbit coupling, and antiferromagnetism in the present case.  When $S$ = 1/2, 3/2, $\cdots$, the fundamental frequency ($p$ = 1) disappears ($R_s$ = 0 for $p$ = 1) but the second harmonic ($p$ =2) remains.  If we assume $S$ = 1/2 and 3/2 near $\theta$ = 20 and 60$^\circ$, respectively, in the $cb$ plane, for example, two spin-zeros occur for the second harmonic between these angles, which explains the suppressed amplitudes of 2$\alpha$ (open pinkish red marks) between the two angles.  Since $m^*$ of the $\alpha$ frequency is almost isotropic in the $ab$ plane (Table I of the journal article), the contrasting behavior of $\alpha$ between the $cb$ and $ca$ planes has to originate from anisotropy of $g_{eff}$; i.e., the observation of spin-zeros only in the $cb$ plane seems to indicate that $g_{eff}$ substantially increases as $B$ approaches $b$, but not as $B$ approaches $a$.   This seems reasonable if we consider the antiferromagnetic order: since the ordered moments are parallel to the $a$ axis, the $b$ axis is the direction of the perpendicular susceptibility of antiferromagnetism, which is not suppressed by the ordering, but the susceptibility along the $a$ axis is suppressed below $T_N$.\\$^1$Shoenberg, D.  \textit{Magnetic oscillations in metals} (Cambridge University Press, Cambridge, 1984)


\begin{thebibliography}{29}
\expandafter\ifx\csname natexlab\endcsname\relax\def\natexlab#1{#1}\fi
\expandafter\ifx\csname bibnamefont\endcsname\relax
  \def\bibnamefont#1{#1}\fi
\expandafter\ifx\csname bibfnamefont\endcsname\relax
  \def\bibfnamefont#1{#1}\fi
\expandafter\ifx\csname citenamefont\endcsname\relax
  \def\citenamefont#1{#1}\fi
\expandafter\ifx\csname url\endcsname\relax
  \def\url#1{\texttt{#1}}\fi
\expandafter\ifx\csname urlprefix\endcsname\relax\def\urlprefix{URL }\fi
\providecommand{\bibinfo}[2]{#2}
\providecommand{\eprint}[2][]{\url{#2}}

\bibitem[{\citenamefont{Kamihara et~al.}(2008)\citenamefont{Kamihara, Watanabe,
  Hirano, and Hosono}}]{Kamihara08JACS}
\bibinfo{author}{\bibfnamefont{Y.}~\bibnamefont{Kamihara}},
  \bibinfo{author}{\bibfnamefont{T.}~\bibnamefont{Watanabe}},
  \bibinfo{author}{\bibfnamefont{M.}~\bibnamefont{Hirano}}, \bibnamefont{and}
  \bibinfo{author}{\bibfnamefont{H.}~\bibnamefont{Hosono}},
  \bibinfo{journal}{J. Am. Chem. Soc.} \textbf{\bibinfo{volume}{130}},
  \bibinfo{pages}{3296} (\bibinfo{year}{2008}).

\bibitem[{\citenamefont{Yin et~al.}(2008)\citenamefont{Yin, Leb\`egue, Han,
  Neal, Savrasov, and Pickett}}]{Yin08PRL}
\bibinfo{author}{\bibfnamefont{Z.~P.} \bibnamefont{Yin}},
  \bibinfo{author}{\bibfnamefont{S.}~\bibnamefont{Leb\`egue}},
  \bibinfo{author}{\bibfnamefont{M.~J.} \bibnamefont{Han}},
  \bibinfo{author}{\bibfnamefont{B.~P.} \bibnamefont{Neal}},
  \bibinfo{author}{\bibfnamefont{S.~Y.} \bibnamefont{Savrasov}},
  \bibnamefont{and} \bibinfo{author}{\bibfnamefont{W.~E.}
  \bibnamefont{Pickett}}, \bibinfo{journal}{Phys. Rev. Lett.}
  \textbf{\bibinfo{volume}{101}}, \bibinfo{pages}{047001}
  (\bibinfo{year}{2008}).

\bibitem[{\citenamefont{Mazin and Johannes}(2009)}]{Mazin09NatPhys}
\bibinfo{author}{\bibfnamefont{I.~I.} \bibnamefont{Mazin}} \bibnamefont{and}
  \bibinfo{author}{\bibfnamefont{M.~D.} \bibnamefont{Johannes}},
  \bibinfo{journal}{Nature Phys.} \textbf{\bibinfo{volume}{5}},
  \bibinfo{pages}{141} (\bibinfo{year}{2009}).

\bibitem[{\citenamefont{Ishibashi and Terakura}(2008)}]{Ishibashi08JPSJS}
\bibinfo{author}{\bibfnamefont{S.}~\bibnamefont{Ishibashi}} \bibnamefont{and}
  \bibinfo{author}{\bibfnamefont{K.}~\bibnamefont{Terakura}},
  \bibinfo{journal}{J. Phys. Soc. Jpn.} \textbf{\bibinfo{volume}{77SC}},
  \bibinfo{pages}{91} (\bibinfo{year}{2008}).

\bibitem[{\citenamefont{Nakamura et~al.}(2008)\citenamefont{Nakamura, Hayashi,
  Nakai, and Machida}}]{Nakamura08condmat}
\bibinfo{author}{\bibfnamefont{H.}~\bibnamefont{Nakamura}},
  \bibinfo{author}{\bibfnamefont{N.}~\bibnamefont{Hayashi}},
  \bibinfo{author}{\bibfnamefont{N.}~\bibnamefont{Nakai}}, \bibnamefont{and}
  \bibinfo{author}{\bibfnamefont{M.}~\bibnamefont{Machida}},
  \bibinfo{journal}{arXiv:0806.4804}  (\bibinfo{year}{2008}).

\bibitem[{\citenamefont{Yin et~al.}(2011)\citenamefont{Yin, Haule, and
  Kotliar}}]{Yin11NatPhys}
\bibinfo{author}{\bibfnamefont{Z.~P.} \bibnamefont{Yin}},
  \bibinfo{author}{\bibfnamefont{K.}~\bibnamefont{Haule}}, \bibnamefont{and}
  \bibinfo{author}{\bibfnamefont{G.}~\bibnamefont{Kotliar}},
  \bibinfo{journal}{Nature Phys.} \textbf{\bibinfo{volume}{7}},
  \bibinfo{pages}{294} (\bibinfo{year}{2011}).

\bibitem[{\citenamefont{Chu et~al.}(2010)\citenamefont{Chu, Analytis, De~Greve,
  McMahon, Islam, Yamamoto, and Fisher}}]{Chu10Science}
\bibinfo{author}{\bibfnamefont{J.-H.} \bibnamefont{Chu}},
  \bibinfo{author}{\bibfnamefont{J.~G.} \bibnamefont{Analytis}},
  \bibinfo{author}{\bibfnamefont{K.}~\bibnamefont{De~Greve}},
  \bibinfo{author}{\bibfnamefont{P.~L.} \bibnamefont{McMahon}},
  \bibinfo{author}{\bibfnamefont{Z.}~\bibnamefont{Islam}},
  \bibinfo{author}{\bibfnamefont{Y.}~\bibnamefont{Yamamoto}}, \bibnamefont{and}
  \bibinfo{author}{\bibfnamefont{I.~R.} \bibnamefont{Fisher}},
  \bibinfo{journal}{Science} \textbf{\bibinfo{volume}{329}},
  \bibinfo{pages}{824} (\bibinfo{year}{2010}).

\bibitem[{\citenamefont{Tanatar et~al.}(2010)\citenamefont{Tanatar, Blomberg,
  Kreyssig, Kim, Ni, Thaler, Bud'ko, Canfield, Goldman, Mazin
  et~al.}}]{Tanatar10PRB}
\bibinfo{author}{\bibfnamefont{M.~A.} \bibnamefont{Tanatar}},
  \bibinfo{author}{\bibfnamefont{E.~C.} \bibnamefont{Blomberg}},
  \bibinfo{author}{\bibfnamefont{A.}~\bibnamefont{Kreyssig}},
  \bibinfo{author}{\bibfnamefont{M.~G.} \bibnamefont{Kim}},
  \bibinfo{author}{\bibfnamefont{N.}~\bibnamefont{Ni}},
  \bibinfo{author}{\bibfnamefont{A.}~\bibnamefont{Thaler}},
  \bibinfo{author}{\bibfnamefont{S.~L.} \bibnamefont{Bud'ko}},
  \bibinfo{author}{\bibfnamefont{P.~C.} \bibnamefont{Canfield}},
  \bibinfo{author}{\bibfnamefont{A.~I.} \bibnamefont{Goldman}},
  \bibinfo{author}{\bibfnamefont{I.~I.} \bibnamefont{Mazin}},
  \bibnamefont{et~al.}, \bibinfo{journal}{Phys. Rev. B}
  \textbf{\bibinfo{volume}{81}}, \bibinfo{pages}{184508}
  (\bibinfo{year}{2010}).

\bibitem[{\citenamefont{Liu et~al.}(2009)\citenamefont{Liu, Kondo, Ni,
  Palczewski, Bostwick, Samolyuk, Khasanov, Shi, Rotenberg, Bud'ko
  et~al.}}]{LiuC09PRL}
\bibinfo{author}{\bibfnamefont{C.}~\bibnamefont{Liu}},
  \bibinfo{author}{\bibfnamefont{T.}~\bibnamefont{Kondo}},
  \bibinfo{author}{\bibfnamefont{N.}~\bibnamefont{Ni}},
  \bibinfo{author}{\bibfnamefont{A.~D.} \bibnamefont{Palczewski}},
  \bibinfo{author}{\bibfnamefont{A.}~\bibnamefont{Bostwick}},
  \bibinfo{author}{\bibfnamefont{G.~D.} \bibnamefont{Samolyuk}},
  \bibinfo{author}{\bibfnamefont{R.}~\bibnamefont{Khasanov}},
  \bibinfo{author}{\bibfnamefont{M.}~\bibnamefont{Shi}},
  \bibinfo{author}{\bibfnamefont{E.}~\bibnamefont{Rotenberg}},
  \bibinfo{author}{\bibfnamefont{S.~L.} \bibnamefont{Bud'ko}},
  \bibnamefont{et~al.}, \bibinfo{journal}{Phys. Rev. Lett.}
  \textbf{\bibinfo{volume}{102}}, \bibinfo{pages}{167004}
  (\bibinfo{year}{2009}).

\bibitem[{\citenamefont{Yang et~al.}(2009)\citenamefont{Yang, Zhang, Ou, Zhao,
  Shen, Zhou, Wei, Chen, Xu, He et~al.}}]{Yang09PRL}
\bibinfo{author}{\bibfnamefont{L.~X.} \bibnamefont{Yang}},
  \bibinfo{author}{\bibfnamefont{Y.}~\bibnamefont{Zhang}},
  \bibinfo{author}{\bibfnamefont{H.~W.} \bibnamefont{Ou}},
  \bibinfo{author}{\bibfnamefont{J.~F.} \bibnamefont{Zhao}},
  \bibinfo{author}{\bibfnamefont{D.~W.} \bibnamefont{Shen}},
  \bibinfo{author}{\bibfnamefont{B.}~\bibnamefont{Zhou}},
  \bibinfo{author}{\bibfnamefont{J.}~\bibnamefont{Wei}},
  \bibinfo{author}{\bibfnamefont{F.}~\bibnamefont{Chen}},
  \bibinfo{author}{\bibfnamefont{M.}~\bibnamefont{Xu}},
  \bibinfo{author}{\bibfnamefont{C.}~\bibnamefont{He}}, \bibnamefont{et~al.},
  \bibinfo{journal}{Phys. Rev. Lett.} \textbf{\bibinfo{volume}{102}},
  \bibinfo{pages}{107002} (\bibinfo{year}{2009}).

\bibitem[{\citenamefont{Malaeb et~al.}(2009)\citenamefont{Malaeb, Yoshida,
  Fujimori, Kubota, Ono, Kihou, Shirage, Kito, Iyo, Eisaki
  et~al.}}]{Malaeb09JPSJ}
\bibinfo{author}{\bibfnamefont{W.}~\bibnamefont{Malaeb}},
  \bibinfo{author}{\bibfnamefont{T.}~\bibnamefont{Yoshida}},
  \bibinfo{author}{\bibfnamefont{A.}~\bibnamefont{Fujimori}},
  \bibinfo{author}{\bibfnamefont{M.}~\bibnamefont{Kubota}},
  \bibinfo{author}{\bibfnamefont{K.}~\bibnamefont{Ono}},
  \bibinfo{author}{\bibfnamefont{K.}~\bibnamefont{Kihou}},
  \bibinfo{author}{\bibfnamefont{P.~M.} \bibnamefont{Shirage}},
  \bibinfo{author}{\bibfnamefont{H.}~\bibnamefont{Kito}},
  \bibinfo{author}{\bibfnamefont{A.}~\bibnamefont{Iyo}},
  \bibinfo{author}{\bibfnamefont{H.}~\bibnamefont{Eisaki}},
  \bibnamefont{et~al.}, \bibinfo{journal}{J. Phys. Soc. Jpn.}
  \textbf{\bibinfo{volume}{78}}, \bibinfo{pages}{123706}
  (\bibinfo{year}{2009}).

\bibitem[{\citenamefont{Richard et~al.}(2010)\citenamefont{Richard, Nakayama,
  Sato, Neupane, Xu, Bowen, Chen, Luo, Wang, Dai et~al.}}]{Richard10PRL}
\bibinfo{author}{\bibfnamefont{P.}~\bibnamefont{Richard}},
  \bibinfo{author}{\bibfnamefont{K.}~\bibnamefont{Nakayama}},
  \bibinfo{author}{\bibfnamefont{T.}~\bibnamefont{Sato}},
  \bibinfo{author}{\bibfnamefont{M.}~\bibnamefont{Neupane}},
  \bibinfo{author}{\bibfnamefont{Y.-M.} \bibnamefont{Xu}},
  \bibinfo{author}{\bibfnamefont{J.~H.} \bibnamefont{Bowen}},
  \bibinfo{author}{\bibfnamefont{G.~F.} \bibnamefont{Chen}},
  \bibinfo{author}{\bibfnamefont{J.~L.} \bibnamefont{Luo}},
  \bibinfo{author}{\bibfnamefont{N.~L.} \bibnamefont{Wang}},
  \bibinfo{author}{\bibfnamefont{X.}~\bibnamefont{Dai}}, \bibnamefont{et~al.},
  \bibinfo{journal}{Phys. Rev. Lett.} \textbf{\bibinfo{volume}{104}},
  \bibinfo{pages}{137001} (\bibinfo{year}{2010}).

\bibitem[{\citenamefont{Yi et~al.}(2011)\citenamefont{Yi, Lu, Chu, Analytis,
  Sorini, Kemper, Mo, Moore, Hashimoto, Lee et~al.}}]{Yi11PNAS}
\bibinfo{author}{\bibfnamefont{M.}~\bibnamefont{Yi}},
  \bibinfo{author}{\bibfnamefont{D.~H.} \bibnamefont{Lu}},
  \bibinfo{author}{\bibfnamefont{J.-H.} \bibnamefont{Chu}},
  \bibinfo{author}{\bibfnamefont{J.~G.} \bibnamefont{Analytis}},
  \bibinfo{author}{\bibfnamefont{A.~P.} \bibnamefont{Sorini}},
  \bibinfo{author}{\bibfnamefont{A.~F.} \bibnamefont{Kemper}},
  \bibinfo{author}{\bibfnamefont{S.-K.} \bibnamefont{Mo}},
  \bibinfo{author}{\bibfnamefont{R.~G.} \bibnamefont{Moore}},
  \bibinfo{author}{\bibfnamefont{M.}~\bibnamefont{Hashimoto}},
  \bibinfo{author}{\bibfnamefont{W.~S.} \bibnamefont{Lee}},
  \bibnamefont{et~al.}, \bibinfo{journal}{PNAS} \textbf{\bibinfo{volume}{108}},
  \bibinfo{pages}{6878} (\bibinfo{year}{2011}).

\bibitem[{\citenamefont{Nakajima et~al.}(2011)\citenamefont{Nakajima, Liang,
  Ishida, Tomioka, Kihou, Lee, Iyo, Eisaki, Kakeshita, Ito
  et~al.}}]{Nakajima11PNAS}
\bibinfo{author}{\bibfnamefont{M.}~\bibnamefont{Nakajima}},
  \bibinfo{author}{\bibfnamefont{T.}~\bibnamefont{Liang}},
  \bibinfo{author}{\bibfnamefont{S.}~\bibnamefont{Ishida}},
  \bibinfo{author}{\bibfnamefont{Y.}~\bibnamefont{Tomioka}},
  \bibinfo{author}{\bibfnamefont{K.}~\bibnamefont{Kihou}},
  \bibinfo{author}{\bibfnamefont{C.~H.} \bibnamefont{Lee}},
  \bibinfo{author}{\bibfnamefont{A.}~\bibnamefont{Iyo}},
  \bibinfo{author}{\bibfnamefont{H.}~\bibnamefont{Eisaki}},
  \bibinfo{author}{\bibfnamefont{T.}~\bibnamefont{Kakeshita}},
  \bibinfo{author}{\bibfnamefont{T.}~\bibnamefont{Ito}}, \bibnamefont{et~al.},
  \bibinfo{journal}{PNAS} \textbf{\bibinfo{volume}{108}},
  \bibinfo{pages}{12238} (\bibinfo{year}{2011}).

\bibitem[{\citenamefont{Rotter et~al.}(2008)\citenamefont{Rotter, Tegel,
  Johrendt, Schellenberg, Hermes, and P\"{o}ttgen}}]{Rotter08PRB}
\bibinfo{author}{\bibfnamefont{M.}~\bibnamefont{Rotter}},
  \bibinfo{author}{\bibfnamefont{M.}~\bibnamefont{Tegel}},
  \bibinfo{author}{\bibfnamefont{D.}~\bibnamefont{Johrendt}},
  \bibinfo{author}{\bibfnamefont{I.}~\bibnamefont{Schellenberg}},
  \bibinfo{author}{\bibfnamefont{W.}~\bibnamefont{Hermes}}, \bibnamefont{and}
  \bibinfo{author}{\bibfnamefont{R.}~\bibnamefont{P\"{o}ttgen}},
  \bibinfo{journal}{Phys. Rev. B} \textbf{\bibinfo{volume}{78}},
  \bibinfo{eid}{020503} (\bibinfo{year}{2008}).

\bibitem[{\citenamefont{Sebastian et~al.}(2008)\citenamefont{Sebastian,
  Gillett, Harrison, Lau, Singh, Mielke, and Lonzarich}}]{Sebastian08JPCM}
\bibinfo{author}{\bibfnamefont{S.~E.} \bibnamefont{Sebastian}},
  \bibinfo{author}{\bibfnamefont{J.}~\bibnamefont{Gillett}},
  \bibinfo{author}{\bibfnamefont{N.}~\bibnamefont{Harrison}},
  \bibinfo{author}{\bibfnamefont{P.~H.~C.} \bibnamefont{Lau}},
  \bibinfo{author}{\bibfnamefont{D.~J.} \bibnamefont{Singh}},
  \bibinfo{author}{\bibfnamefont{C.~H.} \bibnamefont{Mielke}},
  \bibnamefont{and} \bibinfo{author}{\bibfnamefont{G.~G.}
  \bibnamefont{Lonzarich}}, \bibinfo{journal}{J. Phys.: Condens. Matter}
  \textbf{\bibinfo{volume}{20}}, \bibinfo{pages}{422203}
  (\bibinfo{year}{2008}).

\bibitem[{\citenamefont{Analytis et~al.}(2009)\citenamefont{Analytis, McDonald,
  Chu, Riggs, Bangura, Kucharczyk, Johannes, and Fisher}}]{Analytis09PRB}
\bibinfo{author}{\bibfnamefont{J.~G.} \bibnamefont{Analytis}},
  \bibinfo{author}{\bibfnamefont{R.~D.} \bibnamefont{McDonald}},
  \bibinfo{author}{\bibfnamefont{J.-H.} \bibnamefont{Chu}},
  \bibinfo{author}{\bibfnamefont{S.~C.} \bibnamefont{Riggs}},
  \bibinfo{author}{\bibfnamefont{A.~F.} \bibnamefont{Bangura}},
  \bibinfo{author}{\bibfnamefont{C.}~\bibnamefont{Kucharczyk}},
  \bibinfo{author}{\bibfnamefont{M.}~\bibnamefont{Johannes}}, \bibnamefont{and}
  \bibinfo{author}{\bibfnamefont{I.~R.} \bibnamefont{Fisher}},
  \bibinfo{journal}{Phys. Rev. B} \textbf{\bibinfo{volume}{80}},
  \bibinfo{eid}{064507} (\bibinfo{year}{2009}).

\bibitem{SM}
{See Supplemental Material for anisotropic amplitudes of $\alpha$ and spin-splitting effects.}

\bibitem[{\citenamefont{Shoenberg}(1984)}]{Shoenberg84}
\bibinfo{author}{\bibfnamefont{D.}~\bibnamefont{Shoenberg}},
  \emph{\bibinfo{title}{Magnetic oscillations in metals}}
  (\bibinfo{publisher}{Cambridge University Press},
  \bibinfo{address}{Cambridge}, \bibinfo{year}{1984}).

\bibitem[{\citenamefont{Carrington et~al.}(2005)\citenamefont{Carrington,
  Fletcher, Cooper, Taylor, Balicas, Zhigadlo, Kazakov, Karpinski, Charmant,
  and Kortus}}]{Carrington05PRB}
\bibinfo{author}{\bibfnamefont{A.}~\bibnamefont{Carrington}},
  \bibinfo{author}{\bibfnamefont{J.~D.} \bibnamefont{Fletcher}},
  \bibinfo{author}{\bibfnamefont{J.~R.} \bibnamefont{Cooper}},
  \bibinfo{author}{\bibfnamefont{O.~J.} \bibnamefont{Taylor}},
  \bibinfo{author}{\bibfnamefont{L.}~\bibnamefont{Balicas}},
  \bibinfo{author}{\bibfnamefont{N.~D.} \bibnamefont{Zhigadlo}},
  \bibinfo{author}{\bibfnamefont{S.~M.} \bibnamefont{Kazakov}},
  \bibinfo{author}{\bibfnamefont{J.}~\bibnamefont{Karpinski}},
  \bibinfo{author}{\bibfnamefont{J.~P.~H.} \bibnamefont{Charmant}},
  \bibnamefont{and} \bibinfo{author}{\bibfnamefont{J.}~\bibnamefont{Kortus}},
  \bibinfo{journal}{Phys. Rev. B} \textbf{\bibinfo{volume}{72}},
  \bibinfo{pages}{060507} (\bibinfo{year}{2005}).

\bibitem[{\citenamefont{Analytis
  et~al.}(2009{\natexlab{b}})\citenamefont{Analytis, Andrew, Coldea, McCollam,
  Chu, McDonald, Fisher, and Carrington}}]{Analytis09PRL}
\bibinfo{author}{\bibfnamefont{J.~G.} \bibnamefont{Analytis}},
  \bibinfo{author}{\bibfnamefont{C.~M.~J.} \bibnamefont{Andrew}},
  \bibinfo{author}{\bibfnamefont{A.~I.} \bibnamefont{Coldea}},
  \bibinfo{author}{\bibfnamefont{A.}~\bibnamefont{McCollam}},
  \bibinfo{author}{\bibfnamefont{J.-H.} \bibnamefont{Chu}},
  \bibinfo{author}{\bibfnamefont{R.~D.} \bibnamefont{McDonald}},
  \bibinfo{author}{\bibfnamefont{I.~R.} \bibnamefont{Fisher}},
  \bibnamefont{and}
  \bibinfo{author}{\bibfnamefont{A.}~\bibnamefont{Carrington}},
  \bibinfo{journal}{Phys. Rev. Lett.} \textbf{\bibinfo{volume}{103}},
  \bibinfo{eid}{076401} (\bibinfo{year}{2009}{\natexlab{b}}).

\bibitem[{\citenamefont{Terashima et~al.}(2010)\citenamefont{Terashima, Kimata,
  Kurita, Satsukawa, Harada, Hazama, Imai, Sato, Kihou, Lee
  et~al.}}]{Terashima10JPSJ}
\bibinfo{author}{\bibfnamefont{T.}~\bibnamefont{Terashima}},
  \bibinfo{author}{\bibfnamefont{M.}~\bibnamefont{Kimata}},
  \bibinfo{author}{\bibfnamefont{N.}~\bibnamefont{Kurita}},
  \bibinfo{author}{\bibfnamefont{H.}~\bibnamefont{Satsukawa}},
  \bibinfo{author}{\bibfnamefont{A.}~\bibnamefont{Harada}},
  \bibinfo{author}{\bibfnamefont{K.}~\bibnamefont{Hazama}},
  \bibinfo{author}{\bibfnamefont{M.}~\bibnamefont{Imai}},
  \bibinfo{author}{\bibfnamefont{A.}~\bibnamefont{Sato}},
  \bibinfo{author}{\bibfnamefont{K.}~\bibnamefont{Kihou}},
  \bibinfo{author}{\bibfnamefont{C.-H.} \bibnamefont{Lee}},
  \bibnamefont{et~al.}, \bibinfo{journal}{J. Phys. Soc. Jpn.}
  \textbf{\bibinfo{volume}{79}}, \bibinfo{pages}{053702}
  (\bibinfo{year}{2010}).

\bibitem[{\citenamefont{Ran et~al.}(2009)\citenamefont{Ran, Wang, Zhai,
  Vishwanath, and Lee}}]{Ran09PRB}
\bibinfo{author}{\bibfnamefont{Y.}~\bibnamefont{Ran}},
  \bibinfo{author}{\bibfnamefont{F.}~\bibnamefont{Wang}},
  \bibinfo{author}{\bibfnamefont{H.}~\bibnamefont{Zhai}},
  \bibinfo{author}{\bibfnamefont{A.}~\bibnamefont{Vishwanath}},
  \bibnamefont{and} \bibinfo{author}{\bibfnamefont{D.-H.} \bibnamefont{Lee}},
  \bibinfo{journal}{Phys. Rev. B} \textbf{\bibinfo{volume}{79}},
  \bibinfo{pages}{014505} (\bibinfo{year}{2009}).

\bibitem[{\citenamefont{Shimojima et~al.}(2010)\citenamefont{Shimojima,
  Ishizaka, Ishida, Katayama, Ohgushi, Kiss, Okawa, Togashi, Wang, Chen
  et~al.}}]{Shimojima10PRL}
\bibinfo{author}{\bibfnamefont{T.}~\bibnamefont{Shimojima}},
  \bibinfo{author}{\bibfnamefont{K.}~\bibnamefont{Ishizaka}},
  \bibinfo{author}{\bibfnamefont{Y.}~\bibnamefont{Ishida}},
  \bibinfo{author}{\bibfnamefont{N.}~\bibnamefont{Katayama}},
  \bibinfo{author}{\bibfnamefont{K.}~\bibnamefont{Ohgushi}},
  \bibinfo{author}{\bibfnamefont{T.}~\bibnamefont{Kiss}},
  \bibinfo{author}{\bibfnamefont{M.}~\bibnamefont{Okawa}},
  \bibinfo{author}{\bibfnamefont{T.}~\bibnamefont{Togashi}},
  \bibinfo{author}{\bibfnamefont{X.-Y.} \bibnamefont{Wang}},
  \bibinfo{author}{\bibfnamefont{C.-T.} \bibnamefont{Chen}},
  \bibnamefont{et~al.}, \bibinfo{journal}{Phys. Rev. Lett.}
  \textbf{\bibinfo{volume}{104}}, \bibinfo{pages}{057002}
  (\bibinfo{year}{2010}).

\bibitem[{\citenamefont{Rotundu et~al.}(2010)\citenamefont{Rotundu, Freelon,
  Forrest, Wilson, Valdivia, Pinuellas, Kim, Kim, Islam, Bourret-Courchesne
  et~al.}}]{Rotundu10PRB}
\bibinfo{author}{\bibfnamefont{C.~R.} \bibnamefont{Rotundu}},
  \bibinfo{author}{\bibfnamefont{B.}~\bibnamefont{Freelon}},
  \bibinfo{author}{\bibfnamefont{T.~R.} \bibnamefont{Forrest}},
  \bibinfo{author}{\bibfnamefont{S.~D.} \bibnamefont{Wilson}},
  \bibinfo{author}{\bibfnamefont{P.~N.} \bibnamefont{Valdivia}},
  \bibinfo{author}{\bibfnamefont{G.}~\bibnamefont{Pinuellas}},
  \bibinfo{author}{\bibfnamefont{A.}~\bibnamefont{Kim}},
  \bibinfo{author}{\bibfnamefont{J.-W.} \bibnamefont{Kim}},
  \bibinfo{author}{\bibfnamefont{Z.}~\bibnamefont{Islam}},
  \bibinfo{author}{\bibfnamefont{E.}~\bibnamefont{Bourret-Courchesne}},
  \bibnamefont{et~al.}, \bibinfo{journal}{Phys. Rev. B}
  \textbf{\bibinfo{volume}{82}}, \bibinfo{pages}{144525}
  (\bibinfo{year}{2010}).

\bibitem[{\citenamefont{Huang et~al.}(2008)\citenamefont{Huang, Qiu, Bao,
  Green, Lynn, Gasparovic, Wu, Wu, and Chen}}]{Huang08PRL}
\bibinfo{author}{\bibfnamefont{Q.}~\bibnamefont{Huang}},
  \bibinfo{author}{\bibfnamefont{Y.}~\bibnamefont{Qiu}},
  \bibinfo{author}{\bibfnamefont{W.}~\bibnamefont{Bao}},
  \bibinfo{author}{\bibfnamefont{M.~A.} \bibnamefont{Green}},
  \bibinfo{author}{\bibfnamefont{J.~W.} \bibnamefont{Lynn}},
  \bibinfo{author}{\bibfnamefont{Y.~C.} \bibnamefont{Gasparovic}},
  \bibinfo{author}{\bibfnamefont{T.}~\bibnamefont{Wu}},
  \bibinfo{author}{\bibfnamefont{G.}~\bibnamefont{Wu}}, \bibnamefont{and}
  \bibinfo{author}{\bibfnamefont{X.~H.} \bibnamefont{Chen}},
  \bibinfo{journal}{Phys. Rev. Lett.} \textbf{\bibinfo{volume}{101}},
  \bibinfo{pages}{257003} (\bibinfo{year}{2008}).

\bibitem[{\citenamefont{Seah and Dench}(1979)}]{Seah79SIA}
\bibinfo{author}{\bibfnamefont{M.~P.} \bibnamefont{Seah}} \bibnamefont{and}
  \bibinfo{author}{\bibfnamefont{W.~A.} \bibnamefont{Dench}},
  \bibinfo{journal}{Surface and Interface Analysis}
  \textbf{\bibinfo{volume}{1}}, \bibinfo{pages}{2} (\bibinfo{year}{1979}).

\bibitem[{\citenamefont{Huynh et~al.}(2011)\citenamefont{Huynh, Tanabe, and
  Tanigaki}}]{Huynh11PRL}
\bibinfo{author}{\bibfnamefont{K.~K.} \bibnamefont{Huynh}},
  \bibinfo{author}{\bibfnamefont{Y.}~\bibnamefont{Tanabe}}, \bibnamefont{and}
  \bibinfo{author}{\bibfnamefont{K.}~\bibnamefont{Tanigaki}},
  \bibinfo{journal}{Phys. Rev. Lett.} \textbf{\bibinfo{volume}{106}},
  \bibinfo{pages}{217004} (\bibinfo{year}{2011}).

\bibitem[{\citenamefont{Dusza et~al.}(2011)\citenamefont{Dusza, Lucarelli,
  Pfuner, Chu, Fisher, and Degiorgi}}]{Dusza11EPL}
\bibinfo{author}{\bibfnamefont{A.}~\bibnamefont{Dusza}},
  \bibinfo{author}{\bibfnamefont{A.}~\bibnamefont{Lucarelli}},
  \bibinfo{author}{\bibfnamefont{F.}~\bibnamefont{Pfuner}},
  \bibinfo{author}{\bibfnamefont{J.-H.} \bibnamefont{Chu}},
  \bibinfo{author}{\bibfnamefont{I.~R.} \bibnamefont{Fisher}},
  \bibnamefont{and} \bibinfo{author}{\bibfnamefont{L.}~\bibnamefont{Degiorgi}},
  \bibinfo{journal}{EPL (Europhysics Letters)} \textbf{\bibinfo{volume}{93}},
  \bibinfo{pages}{37002} (\bibinfo{year}{2011}).

\end{thebibliography}
\end{document}